\documentclass[a4paper,10pt,twocolumn]{revtex4}
\usepackage[english]{babel}
\usepackage{indentfirst}
\usepackage{graphicx}
\usepackage{amsmath}
\usepackage{amssymb}
\usepackage{amsthm}
\usepackage{mathrsfs}
\usepackage{bbold}
\usepackage{lscape}
\usepackage{bm}
\usepackage{bbm}
\usepackage{float}
\usepackage{longtable}
\usepackage{graphics}
\usepackage{color}
\usepackage[utf8x]{inputenc}

\def\ket#1{\left|#1\right>}
\def\bra#1{\left<#1\right|}

\begin{document}

\title{Nuclear spin cooling using Overhauser field selective coherent population trapping}
\author{M. Issler$^{1,*}$ \footnote[0]{\scriptsize$^{\rm *}$ These authors have contributed equally to this work.},
E. Kessler$^{2,*}$, G. Giedke$^2$, S. Yelin$^3$, I. Cirac$^2$, M. Lukin$^4$, A.
Imamoglu$^1$} \affiliation{$^1$ Institute of
  Quantum Electronics, ETH-Z\"urich, CH-8093 Z\"urich, Switzerland}
\affiliation{$^2$ Max-Planck-Institut für Quantenoptik, Hans-Kopfermann-Str. 1
85748 Garching,
  Germany}
\affiliation{$^3$ Department of Physics, University of Connecticut 2152 Hillside
Road, U-3046 Storrs, CT 06269-3046, USA} \affiliation{$^4$ Department of Physics,
Harvard University, Cambridge, MA 02138, USA}

\vspace{-3.5cm}

\date{\today}
\begin{abstract}
Hyperfine interactions with a nuclear spin environment fundamentally limit the
coherence properties of confined electron spins in the solid-state. Here, we show
that a quantum interference effect in optical absorption from two electronic spin
states of a solid-state emitter  can be used to prepare the surrounding
environment of nuclear spins in  well-defined states, thereby suppressing
electronic spin dephasing.  The evolution of the coupled electron-nuclei system
into a coherent population trapping state by optical excitation induced nuclear
spin diffusion can be described in terms of L\'evy flights, in close analogy with
sub-recoil laser cooling of atoms. The large difference in electronic and nuclear
time scales simultaneously allow for a measurement of the magnetic field produced
by nuclear spins, making it possible to turn the lasers that cause the anomalous
spin diffusion process off when the strength of the resonance fluorescence reveals
that the nuclear spins are in the desired narrow state.
\end{abstract} \pacs{} \maketitle

The phenomenon of coherent population trapping (CPT) in three-level emitters
\cite{CPT} is at the heart of a number of key advances in quantum optics, such as
sub-recoil cooling of atoms \cite{Aspect88} and slow-light propagation
\cite{Hau99,Lukin00,EIT05}. In these experiments, optical excitation from two low
energy (spin) states to a common optically excited state vanishes due to a quantum
interference effect, leading to the formation of a {\sl dark resonance} whenever
the two driving laser fields satisfy the two-photon resonance condition. The
fundamental limit on how well quantum interference eliminates optical absorption
is provided by the decoherence rate of the two low-energy spin states. Typically,
this decoherence rate is assumed to be induced by a reservoir which could be
treated using the usual Born-Markov approximation, implying that the reservoir has
a short correlation time and its density operator is not influenced by the
interactions.

Unlike their atomic counterparts, solid-state spins are in general
subject to non-Markovian dephasing \cite{Loss02,Taylor07,Claassen10}
due to their coupling to reservoirs with long correlation times. In
particular, hyperfine coupling to nuclear spins constitutes the most
important source of decoherence for spin qubits. It has been
proposed that polarizing or cooling nuclear spins could alleviate
this decoherence process \cite{Loss02}, which prompted theoretical
\cite{Giedke06,Klauser06} as well as experimental efforts aimed at
narrowing down the Overhauser field distribution
\cite{Reilly08,Vink09,Latta09}. These schemes could be considered as
a form of {\sl reservoir engineering}; remarkably, recent
experiments showed that the substantial manipulation of the nuclear
spins (reservoir) could be achieved by using the electron spin
(system) itself
\cite{Brown96,Eble06,Maletinsky07,Vink09,Latta09,Chekhovich10}.

In this Letter, we show that CPT in the spin states of a solid-state
emitter could be used to prepare a nuclear spin environment in
states with a near-deterministic Overhauser field. The preparation
of an ultra-narrow nuclear spin distribution is achieved by optical
excitation induced anomalous diffusion processes \cite{Aspect88}. As
a consequence of the anomalous diffusion, the coupled
electron-nuclei system dynamically switches back and forth between a
trapped regime where nuclear spin diffusion slows down drastically
due to the formation of a dark state, and a non-trapped regime where
optical excitation leads to fast diffusion \cite{Bardou94}. We find
that for a range of system parameters, the coupled system finds the
dark state via this diffusion process and then remains trapped in it
for long times, ensuring a narrow nuclear spin distribution with a
standard deviation that is close to the single-spin limit. An
additional remarkable feature of the scheme that we analyze is the
possibility of using resonantly scattered photons to measure which
regime the coupled system is in at a given time \cite{Stepanenko06};
turning the laser fields off after determining the coupled system to
be in the dark state can then be used to further narrow down the
Overhauser field distribution to the sub-single-spin regime. Such a
feedback mechanism is enabled by the large difference between the
time scales corresponding to electronic light scattering and nuclear
spin-flip processes.

\vspace{0.5 cm}

{\bf Nuclear-spin selective coherent population trapping}

We consider a solid-state emitter where the two ground electronic
spin states, denoted by $\ket{\uparrow_x} $ and $\ket{\downarrow_x}
$, are coupled by two laser fields to a common optically excited
state $\ket{t}$ (Fig.~1a). The laser field with frequency $\omega_p$
($\omega_c$) that couples the $\ket{\uparrow_x} - \ket{t} $
($\ket{\downarrow_x} - \ket{t} $) transition with Rabi frequency
$\Omega_p$ ($\Omega_c$) is referred to as the probe (coupling)
field. The state $\ket{t} $ decays in turn via spontaneous emission
back to the two ground spin states with an equal rate $\Gamma_{t
\uparrow} = \Gamma_{t \downarrow} = \Gamma/2$. Denoting the Zeeman
energy of the electron spin due to the external field $B_x$ with
$\omega_x$ and the energy of the optically excited state with
$\omega_t$, we express the bare optical detunings relevant for the
CPT system as $\Delta\omega_p = \omega_t - \omega_p$ and
$\Delta\omega_c = \omega_t - \omega_x - \omega_c$. In the absence of
any spin interactions or decoherence, laser fields satisfying the
two-photon resonance condition ($\delta = \Delta\omega_p
-\Delta\omega_c = 0$) pump the electron spin into the dark state
$|D\rangle = \frac{\Omega_c}{\sqrt{\Omega_p^2 + \Omega_c^2}} \,
\ket{\uparrow_x} - \frac{\Omega_p}{\sqrt{\Omega_p^2 + \Omega_c^2}}
\, \ket{\downarrow_x},$ which is decoupled from optical excitation.
When $\Delta\omega_c = 0$ and $\Omega_p , \Omega_c \ll \Gamma$, the
absorption lineshape of the emitter appears as a Lorentzian with a
quantum interference induced transparency dip in the center, with a
width $\delta \nu_\mathrm{trans} \sim (\Omega_p^2 +
\Omega_c^2)/\Gamma \ll \Gamma$.

In practice, the electronic spin states of most solid-state emitters
are mutually coupled via hyperfine interaction with a nuclear spin
ensemble consisting of $N$ nuclei
\begin{equation}
\label{eq:hyp}
   H_{\mathrm{hyp}} = g\sum_i^N g_i \left( I_x^i \sigma_x
              + \frac{1}{2}(I_+^i \sigma_- + I_-^i \sigma_+)\right)
              \; .
\end{equation}
Here, $g_i$ defines the normalized hyperfine coupling constant between the emitter
electron and the $i^{th}$ nucleus ($\sum g_i^2=1$). In this convention $g=A_H/\sum
g_i$ quantifies the collective hyperfine coupling strength, with $A_H$ denoting
the hyperfine interaction constant of the material. $\sigma_k$ and $I_\alpha^i$
($\alpha=+,-,x$) are the electronic and nuclear spin operators, respectively;
$\sigma_+  = \ket{\uparrow_x}  \bra{\downarrow_x}$.

Our analysis of CPT in the presence of hyperfine interactions with a nuclear spin
reservoir starts with  the  master equation, obtained by eliminating the radiation
field reservoir using a Born-Markov approximation:
\begin{equation}
\label{eq:mast}
    \dot{\rho} = \frac{\Gamma}{2} ( \mathbb{1}_S \otimes \rho_{tt}  - \{ |t\rangle \langle t| ,\rho \}_+)
    -  i [H_0 + H_\mathrm{laser} + H_\mathrm{hyp},\rho],
\end{equation}
where $\rho_{tt}=\langle t|\rho|t \rangle$ acts on the Hilbert space
of nuclear spins and $ \mathbb{1}_S =
\ket{\uparrow_x}\bra{\uparrow_x}
+\ket{\downarrow_x}\bra{\downarrow_x} $. We assume that in the
absence of optical excitation, the electron spin is well isolated
from all reservoirs other than the nuclear spins \cite{Atature06},
and spin-flip co-tunneling or phonon emission rates are negligible
within the timescales of interest.

In the limit of a large external field ($\omega_x \gg g$), the
direct electron-nuclei flip-flop processes $I_+ \sigma_- + I_-
\sigma_+$ (collective spin operators are defined as
$I_{\alpha}=\sum_i g_i I_\alpha^i$) are strongly suppressed due to
the large mismatch in the electronic and nuclear Zeeman splitting.
In contrast, optical excitation does allow for energy conservation
in an optically assisted electron-nuclear spin-flip process. We take
the higher order processes into account by applying a
Schrieffer--Wolff transformation to eliminate the direct hyperfine
flip-flop interaction. The master equation  then reads
\begin{align}
\label{eq:SW}
    \dot{\rho}=&\frac{\Gamma}{2} ( \mathbb{1}_S \otimes \rho_{tt}  - \{ |t \rangle \langle t|,\rho \}_+)
    - i [H_0 + H_\mathrm{laser} + \tilde{H}_\mathrm{hyp},\rho] \nonumber \\
        &+ \epsilon^2 \frac{\Gamma}{4} \mathbb{1}_S\otimes D(\rho_{tt}) \\
        =& \mathcal{L}_0 (\rho) + \epsilon^2 \mathcal{L}_1(\rho_{tt}) \nonumber,
\end{align}
where the new term containing
\begin{align}
\label{eq:D} D(\rho)=I_+ \rho I_- + I_- \rho I_+  - \frac{1}{2}
\{I_+I_- +I_-I_+,\rho \}_+
\end{align}
describes an optically induced random nuclear diffusion process
caused by the optically assisted hyperfine flip-flop processes, that
are lowest order in the parameter $\epsilon = g/(2 \omega_x)$. In
Eq.~(\ref{eq:SW}) we have neglected terms $\propto \epsilon^2$ that
only affect the electron evolution \cite{SW-note}.

After the Schrieffer--Wolff transformation the Hamiltonian relevant
for electron spin dynamics (to highest order in $\epsilon$) is
$\tilde{H}_{\mathrm{spin}} = \tilde{H}_\mathrm{hyp} + \delta
\sigma_x = g \sigma_x (I_x + \epsilon I_+ I_-+\delta/g)$. The
electron experiences an effective magnetic field, which is composed
of the two-photon detuning $\delta$, as well as a contribution
originating from the nuclei, which we refer to as the generalized
Overhauser field $\tilde{I}_x=I_x + \epsilon I_+ I_-$. For a given
laser detuning $\delta$  each eigenvector of the generalized
Overhauser field $\tilde{I}_x\ket{\lambda}=\lambda \ket{\lambda}$
corresponds to a steady state $\rho_{\lambda}=\rho^e(\lambda)
\otimes \ket{\lambda}\bra{\lambda}$ of the unperturbed evolution
$\mathcal{L}_0(\rho_{\lambda})=0$. Here, $\rho^e(\lambda)$ is given
as the solution of the optical Bloch equations (OBE) found after
projection of the unperturbed master equation on the respective
nuclear state $\ket{\lambda}$; in the OBE
$\delta_\mathrm{eff}=g\lambda + \delta$ gives the effective
two-photon detuning that determines the CPT condition. The lifetime
of such quasi-steady states $\rho_{\lambda}$ under the full dynamics
of Eq.~(\ref{eq:SW}) is determined by hyperfine assisted scattering
events, which are described by the term $\mathcal{L}_1$. The
corresponding nuclear spin flip rate in positive (negative)
direction $D^+$ ($D^-$) can be directly deduced from
Eq.~(\ref{eq:SW}):
$D^\pm=\epsilon^2\frac{\Gamma}{2}\rho^e_{tt}(\lambda) \left< I_\mp
I_\pm\right>$, where
$\rho^e_{tt}(\lambda)=\bra{t}\rho^e(\lambda)\ket{t}$ is the
population in state $\ket{t}$. Each nuclear spin flip event of this
kind changes $\langle\tilde{I}_x\rangle$ by a value of order $g_i$.

For nuclear states with $g\lambda = -\delta$ the system is in
two-photon resonance and the electronic system is transparent such
that $\rho^e_{tt} = 0$: as a consequence, the nuclear spin diffusion
vanishes and the system is trapped in a dark state.  Since the
generalized Overhauser field in an electronic--nuclear dark state is
locked to a fixed value, its variance will be strongly reduced
(nuclear state narrowing) suppressing hyperfine-induced electron
spin decoherence. Strikingly, by narrowing the generalized
Overhauser field, even electron-mediated nuclear spin diffusion is
suppressed, thus eliminating the second order contribution to
hyperfine-induced electron spin decoherence as well. For all nuclear
states satisfying $g\lambda \approx -\delta$, the excited electronic
state population will remain small ($\rho^e_{tt} \propto
\delta_\mathrm{eff}^2$), ensuring that the spin diffusion rate will
remain vanishingly small: we refer to this subspace as the trapping
region.

In contrast, nuclear states with $g\lambda \not\approx -\delta$
render the electron optically active and the generalized Overhauser
field experiences random diffusion (recycling region).  To
illustrate the dynamics allowing the nuclei to move from the
recycling to the trapping region, we consider an electron that is
optically excited to state $\ket{t}$: as it decays, it could induce
a nuclear spin flip event (with probability $\sim \epsilon^2$) in
either direction. Through successive spin-flip events, the nuclear
reservoir probes different spin configurations with distinct
generalized Overhauser shifts. When the diffusion allows the nuclei
to reach a configuration that yields $\delta_\mathrm{eff} \approx
0$, the electron becomes trapped in the dark state; further optical
excitation is then inhibited and nuclear spin flips are strongly
suppressed.

Owing to the quasi-continuous nature of the generalized Overhauser
field spectrum, the dark-state condition $\delta_\mathrm{eff} = 0$
can be satisfied for a wide range of initial detunings
$\Delta\omega_p$.  This leads to a drastic change in the CPT
signature in absorption spectroscopy: instead of exhibiting a narrow
transparency dip at (bare) two-photon-resonance ($\delta=0$), the
coupled electron-nuclei system displays a broad transparency window.

The operator valued correction $\tilde{I}_x$ to the two-photon
detuning $\delta$ and the optically induced diffusive dynamics of
$\tilde{I}_x$ described by the second line of Eq.~(\ref{eq:SW}) are
at the heart of the nuclear-spin cooling scheme we analyze in this
work. The predictions we outlined hold in general for any nuclear
operator $\tilde{I}_x$ with a sufficiently large density of states
around $g\lambda+\delta=0$. In the Methods we show that this
requirement is fulfilled for the generalized Overhauser field
$\tilde{I}_x = I_x +\epsilon I_+ I_-$ and that its properties are
very similar to those of $I_x$ for the parameters we consider.
Therefore, for the sake of simplicity, we will proceed by neglecting
the $\epsilon$ correction. As a further simplification we will
constrain our analysis to nuclear spin $1/2$ systems. While our
results apply to a broad class of solid-state emitters, ranging from
various types of quantum dots to NV centers, we will focus primarily
on a single electron charged quantum dot (QD) where the optically
excited state is a trion state consisting of an electron singlet and
a valence-band hole (Fig.~1a)
\cite{Stepanenko06,Steel09,Warburton09}. For most QD systems, the
assumptions we stated earlier are realized in Voigt geometry where
$B_x$ is applied perpendicular to the growth direction.

\vspace{0.5 cm}

{\bf Semiclassical analysis}

We first consider the semiclassical limit to numerically confirm the
principal striking features of the coupled electron-nuclei system --
altered CPT signatures and the drastic nuclear state narrowing --
for inhomogeneous electron-nuclear coupling. To obtain a
semiclassical description of the coupled electron-nuclei dynamics,
we start by assuming that the electron ($\rho^e$) and the nuclear
($\rho^n$) spins remain unentangled throughout the system evolution
($\rho = \rho^e \otimes \rho^n$). Since, as discussed earlier, the
electron dynamics takes place on a timescale that is faster by a
factor $ \epsilon^{-2} \gg 1$ than the nuclear dynamics, it is
justified to solve the OBE in steady state to determine the trion
population $\rho^e_{tt} $ for a given nuclear spin configuration
(and the associated effective magnetic field).

To describe the nuclear spin dynamics semiclassically, we assume
that the nuclear density operator $\rho^n$ is diagonal in the basis
of individual nuclear spin eigenstates. This assumption is justified
for QDs in which either strongly inhomogeneous hyperfine coupling or
inhomogeneous quadrupolar fields lead to large variations in the
splitting of the nuclear spin states; when this is the case, the
nuclear superposition states will effectively dephase, justifying
the assumption of a diagonal density operator. In this limit, the
master equation Eq.~(\ref{eq:SW}) reduces to rate equations which
can be numerically solved using Monte Carlo techniques (see
Methods).

Figure~2 shows the result of the Monte Carlo simulations of the
coupled electron-nuclei evolution. To obtain the probe field
absorption lineshape as well as the Overhauser field variance, we
assume that for each probe field detuning, we start out from a
completely mixed $\rho^n$, take $\Delta \omega_c = 0$ and evolve the
coupled system to its steady state for a range of probe laser
detunings. We find that the transparency window that has a width of
$\sim 0.12 \Gamma$ ($ \sim 0.48 \Gamma$) for $\Omega_c = \Omega_p =
0.2 \Gamma$ ($\Omega_c = \Omega_p = 0.4 \Gamma$) in the absence of
hyperfine coupling (Fig.~2a, red dashed curve) is drastically
broadened and assumes a width $\delta \nu_\mathrm{trans}
> \Gamma$ (Fig.~2a, solid curves). This {\sl dragging of
the dark resonance} effect is in contrast to Faraday geometry
experiments where nuclear spin polarization ensures that the applied
laser field remains locked to a detuning that ensures maximal
absorption \cite{Latta09}. Concurrently, the Overhauser field
distribution is narrowed dramatically from its value in the absence
of optical excitation (Fig.~2b, black dashed line) such that its
standard deviation $\sigma_{OF}$ is smaller than the change induced
by flipping one nuclear spin of the most weakly coupled class
(Fig.~2b, solid curves). These simulations show all the striking
features that are a consequence of the optically induced nuclear
spin diffusion [second line of Eq.~(\ref{eq:SW})] which leads to a
uni-directional evolution into the electronic-nuclear dark state
$\rho_D = |D\rangle \langle D| \otimes \rho^n_D$, where $\rho^n_D$
is a nuclear spin density operator that yields
$\delta_{\mathrm{eff}} = 0$.

We remark that the narrowing of the Overhauser field distribution
could be measured by using the same two-laser set-up and scanning
the probe laser on timescales short compared to those required to
polarize the nuclear spins, thanks to the large separation between
the electronic and nuclear dynamical timescales. Figure~2c shows the
simulation of the absorption lineshape obtained by such probe laser
scans. Starting out with 100 random nuclear spin configurations, we
first let the coupled electron-nuclei system evolve to its
steady-state under initial laser detunings $\Delta \omega_c = 0$ and
$\Delta \omega_p = -0.2\Gamma$. We then scan the probe laser in
either direction to obtain the absorption lineshape (solid blue
curve) that directly reveals information about the narrowing of the
Overhauser field distribution. If we repeat the numerical experiment
by assuming that the laser fields were initially completely off
resonance, we find that the absorption lineshape is nearly
Lorentzian (dashed green curve).

\vspace{0.5 cm}

{\bf Quantum model of the nuclear spin dynamics}

Next, we study the homogeneous coupling limit using a full quantum treatment. To
capture the full quantum dynamics, we derive a master equation which depends only
on nuclear degrees of freedom, allowing for both an analytical steady state
solution and the comparison between the quantum and the semiclassical limit. To
this end, we assume the homogeneous limit ($g_i=1/\sqrt{N}$). First, we eliminate
the state $|t\rangle$ in the limit $\Omega_p, \Omega_c\ll \Gamma$, giving a master
equation involving the nuclear and electronic spins only.  We also assume
$\Omega_c=\Omega_p = \Omega$, which ensures that the relevant electron spin states
in the rotating frame are $|D \rangle =|\!\downarrow_z\rangle $ and $|B \rangle
=|\!\uparrow_z \rangle$, and choose $\delta=0$, for simplicity. In the interaction
picture we then obtain from Eq.~(\ref{eq:SW}) the reduced master equation
\begin{align}
\label{eq:fin} \dot{\rho}=& \Gamma_\mathrm{eff}(\sigma^z_- \rho \sigma^z_+
- \frac{1}{2} \{\sigma^z_+\sigma^z_-,\rho \}_+) \\\nonumber &+
\frac{\Gamma_\mathrm{eff}}{2} [\sigma_z,[\sigma_z,\rho]]  -i g
I_x [\sigma_x, \rho]\\\nonumber &+\epsilon^2  \mathbb{1}_S\otimes
D(\Gamma_\mathrm{eff} \rho_{\uparrow_z \uparrow_z}),
\end{align}
where $\sigma^z_\pm$ are the electron spin matrices in the $z$-basis and
$\Gamma_\mathrm{eff}=\frac{\Omega^2}{(\Gamma/2)^2 +
  (gI_x/2)^2}\frac{\Gamma}{2}$ is an operator valued effective (electron) spin decay
rate. The last line of Eq.~(\ref{eq:fin}) describes the nuclear spin
diffusion determined by the nuclear operator $\rho_{\uparrow_z
\uparrow_z} = \langle \uparrow_z | \rho | \uparrow_z \rangle$, cf.
Eq.~(\ref{eq:D}) \cite{densitymatrix}.

In order to eliminate the electronic degrees of freedom from
Eq.~(\ref{eq:fin}) we once again make use of the fact that on the timescales
of the electron evolution, the nuclear field can be considered as quasi-static
and hence the electron settles quickly (on nuclear timescales) to its interim
steady state. We find that on this coarse grained timescale $\rho_{\uparrow_z
  \uparrow_z} =
\frac{1}{2}[1-(\frac{\Gamma_\mathrm{eff}}{|\Delta_\mathrm{eff}|})^2]
\rho_n$, with $|\Delta_\mathrm{eff}|^2=\Gamma_\mathrm{eff}^2 + (g
I_x)^2$. Using this relation, the electron spin can be eliminated
from Eq.~(\ref{eq:fin}), yielding
\begin{align}
\label{eq:end} \dot{\rho^n}=&Tr_S(\dot{\rho})=  D(
\Gamma_\mathrm{nuc}  \rho^n),
\end{align}
where we defined the nuclear spin flip rate $\Gamma_\mathrm{nuc} =
\epsilon^2 [1-(\frac{\Gamma_\mathrm{eff}}{|\Delta_\mathrm{eff}|})^2]
\Gamma_\mathrm{eff}$. Note that since $\Gamma_\mathrm{nuc}$ vanishes
for all zero eigenstates of $I_x$, Eq.~(\ref{eq:end}) implies, in
accordance with the considerations above, that every state in the
kernel of the collective nuclear spin operator $I_x$ -- i.e. a state
of vanishing Overhauser field -- is a steady state of the dynamics.
We plot the nuclear spin flip rate in Fig.~3a.

Equation~(\ref{eq:end}) can be used to directly compare the quantum
mechanical and semiclassical diffusion rates in the homogeneous
limit (see Methods). Surprisingly, the two opposite regimes of
semiclassical and quantum mechanical description show both
qualitative (evolution can be fully characterized by rate equations)
and quantitative (for the relevant states the calculated rates are
comparable) agreement (cf. Fig.~3a); this result is particularly
interesting since we would expect the semiclassical description to
fail in the homogeneous limit.

In order to calculate the achievable Overhauser field standard
deviation $\sigma_{OF}$ we numerically compute the exact steady
state solution of master equation~(\ref{eq:mast}) for homogeneously
coupled nuclei. To this end, we explicitly consider all orders of
the hyperfine interaction including processes that result in a
(small) finite decay rate out of the dark state \cite{off-resonant}.
Figure~3b shows $\sigma_{OF}$ as a function of $\Omega$, where we
find that $\sigma_{OF}$ decreases with decreasing $\Omega$ until it
reaches a minimum of $\sigma_{OF} \simeq 2$ ($\sigma_{OF} \simeq
0.7$) for $ \Omega \simeq 0.2 \Gamma$ and an electron spin
decoherence rate of $T_2^{-1} = 100 s^{-1}$ ($T_2^{-1} = 0$). This
result can be understood by recalling that the width of the
transparency dip in CPT scales as $\Omega^2/\Gamma$, implying that
the range of Overhauser field values yielding transparency can be
narrowed simply by reducing $\Omega$. For $\Omega < 0.2 \Gamma$, we
find that $\sigma_{OF}$ increases rapidly; for such small values of
$\Omega$, the coupled electron-nuclei system spends substantial
amount of time outside the narrow transparency region, leading to
the observed increase in steady state value of $\sigma_{OF}$. As we
will argue below, this increase does not constitute a fundamental
limitation for the attainable narrowing and values $\sigma_{OF} < 1$
are possible by using a feedback mechanism. Clearly though, such a
remarkable level of narrowing could only be observed if it is
achieved on timescales short compared to those imposed by electron
spin decoherence and optical excitation independent nuclear spin
decay processes. We now turn to the question \emph{how quickly} the
nuclear spins reach this narrowed state.

{\bf Evolution of the nuclear spins as L\'evy flights}

There is close analogy between the problem of CPT in the presence of inhomogeneous
hyperfine interactions with a slow nuclear spin ensemble and that of a
one-dimensional velocity selective CPT \cite{Aspect88,Bardou94}; the role of
atomic momentum in the latter case is assumed by the nuclear Overhauser field
$I_x$ in the present problem. Just like the atomic momentum along the direction of
interest could change by any value up to the full recoil momentum upon light
scattering, the Overhauser field could change by any value, thanks to an
inhomogeneous distribution of hyperfine interaction constants $0 \le g_i  \le
g_i^{max}$. The two models differ in two important aspects: first, there is a
maximum value of the Overhauser field $\langle g I_x \rangle = A_H$ given by full
polarization of the nuclei, and second, only a small fraction $\epsilon^2 \ll 1$
of light scattering events give rise to a change in the nuclear spin
configuration.

It is known from the velocity selective CPT problem that the
timescales for sub-recoil cooling of the atomic momentum
distribution could be understood using L\'evy flight analysis
\cite{Bardou94}. We apply this method to determine the timescale
over which we expect the nuclear spins to reach a configuration with
a mean Overhauser field that is smaller than a prescribed value
(trapping region). Consider the random walk of the Overhauser field
in time for $\Delta\omega_p = \Delta\omega_c = 0$, $\Omega_p =
\Omega_c = \Omega$: this random walk is characterized by periods of
diffusion followed by long intervals where the Overhauser field is
restricted (trapped) at a value close to $\langle I_x \rangle = 0$.
The duration of the longest trapping interval is typically on the
order of the interaction time  -- a signature of L\'evy statistics.
The probability distribution functions $P(t)$ for the trapping time
$t$, and $\hat P(\hat t)$ for the recyling time $\hat{t}$ for which
the Overhauser field diffuses within the recycling region before
returning to the trap characterize the L\'evy flights
\cite{Bardou94}.

We are mainly interested in  the time required for an Overhauser
field initially in the recycling region to diffuse to a value that
is within a prescribed interval that defines the trapping region.
Once again, this simplification is a consequence of the fact that a
drop in scattered light intensity could reveal whether or not the
nuclear spin distribution has the prescribed value on timescales
smaller by $\epsilon^2$ than those needed to flip another nuclear
spin. A feedback mechanism could therefore ensure that laser
excitation is turned off and the desired/attained $\sigma_{OF}$ is
preserved.  To simplify the estimation of the trapping time, we
consider a limiting case where $ \Omega^2/\Gamma < A_H/N$; i.e. a
typical single nuclear spin flip will take the system out of the
transparency window.

Around the transparency point, the rate at which nuclear spins flip
is given by $D^\pm \propto \langle I_x \rangle^2$. This dependence
yields $P(t) \propto t^{-\frac{3}{2}}$, which in turn leads to
infinite average trapping times \cite{Bardou94}. If we assume that
the width of the recycling region is determined by $
\frac{A_H}{\sqrt{N}} \approx \frac{\Gamma}{4}$, then the light
scattering rate outside the transparency window could be taken to be
constant with value $\Omega^2/\Gamma$ \cite{scat-rate}. The nuclear
spin flip rate in the recycling region is then given by $\tau_0^{-1}
\approx \epsilon^2 \Omega^2/\Gamma$.

In this simplified model the random walk is confined and unbiased so in the limit
of many nuclear spin flips, the number of steps required to return to the trap is
given by $\langle M \rangle
 = \frac{A_H/\sqrt{N}}{\Omega^2/\Gamma}
\approx \frac{A_H/\sqrt{N}}{A_H/N} = \sqrt{N}$. Since the time for a
single spin flip is taken to be independent of the Overhauser field,
the time to return to the trap is given by
\begin{equation}
 \langle \hat t \rangle = \langle M \rangle \tau_0  = \frac{A_H/\sqrt{N}}{\Omega²/\Gamma} \frac{\Gamma}{\Omega^2} \frac{1}{
 \epsilon^2}
\approx \frac{N^{3/2}}{A_H \epsilon^2}.
\end{equation}
For $\omega_x \approx A_H$ this expression simplifies to $\langle
\hat t \rangle \approx N^{\frac{5}{2}}/A_H$.

Given the strong $N$ dependence of $\langle \hat t \rangle$
corresponding to the timescale needed to establish $\sigma_{OF} \sim
1$, it is important to consider nuclear spin dynamics arising from
optical-excitation-independent nuclear spin diffusion or decay
processes, as well as the electron spin decoherence. The ultimate
limit for the latter is due to spin-orbit mediated spin-flip phonon
emission with a rate $\sim 10^{-7} \Gamma$ for $\omega_x \approx
A_H$ \cite{Finley04}; as seen in Fig.~3b the resulting increase in
$\sigma_{OF}$ is a factor $\sim 3$ as compared to the case with no
electron spin decay. Physical processes leading to nuclear spin
diffusion in the dark state include (a) nuclear spin diffusion
mediated by exchange coupling of the QD electron spin to a
degenerate electron gas or by phonon emission/absorption
\cite{Claassen10}, (b) electric field fluctuations in the QD
environment leading to spatial shifts in the electron wave-function,
(c) nuclear quadrupolar fields with axes not parallel to $B_x$. If
we denote the optical excitation independent single nuclear spin
diffusion rate that can arise from any of these mechanisms with
$\gamma_n$ and assume that $N \gamma_n \ll \Omega^2/\Gamma$, then we
could write the steady-state standard deviation of the Overhauser
field as
\begin{equation}
\sigma_{OF} \simeq \tilde{\delta} \frac{\langle t\rangle}{\langle
t\rangle + \langle \hat t \rangle} + \frac{A_H}{\sqrt{N}}
\frac{\langle \hat t\rangle}{\langle t\rangle + \langle \hat t
\rangle}
\end{equation}
where the average time spent in the trapping region $\langle
t\rangle = (N \gamma_n)^{-1}$ and the effective width of the trap
$\tilde{\delta} = \epsilon^{-1} \Omega \sqrt{N \gamma_n/\Gamma}$.
The optimal condition for measurement independent Overhauser field
narrowing is obtained when the two contributions to $\sigma_{OF}$
are comparable. As we have argued earlier, the use of feedback from
the resonance fluorescence intensity should allow for reaching
$\sigma_{OF} \sim \tilde{\delta}$.

\vspace{0.5 cm}

{\bf Prospects for experimental realization}

We have seen that the strength $\gamma_n$ of optical excitation
independent nuclear spin diffusion processes determines the degree
of attainable Overhauser field narrowing. In this context, we remark
that experimental observations reported by the Bayer group
\cite{Greilich07}, obtained by driving an ensemble of
single-electron charged QDs using periodic ultra-short optical
pulses in the Voigt geometry, demonstrated that optically prepared
nuclear spin states could survive for $\sim 10$ minutes
\cite{Bayer-desc}. Such long nuclear spin lifetimes in principle
allow for reaching $\sigma_{OF} \sim 1$ using the proposed CPT
scheme. We also note that the basic signatures of CPT have been
observed in both single electron \cite{Steel09} and hole
\cite{Warburton09} charged QDs.

Even though we have concentrated on nuclear spin diffusion associated with the
ground-state hyperfine coupling, the conclusions of our work remain unchanged if
the solid-state emitter has hyperfine coupling leading to nuclear spin diffusion
in the optically excited state. This would be the case for example in QDs with
vanishing heavy-light hole mixing leading to near-resonant hole-mediated nuclear
spin-flips in the excited state due to the dominant $S_z^h I_z$ term in the
hole-hyperfine interaction Hamiltonian.

While prior experimental results on pulsed excitation of an ensemble
of QDs strongly suggest the feasibility of our proposal in
self-assembled QDs, we expect our findings to be relevant for a
wider range of solid-state emitters. Of particular interest is
nitrogen-vacancy (NV) centers in diamond where CPT has also been
previously observed \cite{Santori06}. The small number of nuclear
spins coupled to the optically excited spin in the case of
NV-centers should make it possible to reduce the time needed for the
system to find the dark state drastically. A principal difference
with respect to the large $N$ limit we analyzed is the fact that
only a small set of optical detunings will allow the NV system to
find a dark state. Finally, extensions to other solid state systems
such as superconducting qubits may be possible \cite{Kelly10}.

\begin{acknowledgments}
  We acknowledge support by an ERC Advanced Investigator Grant (M.I. and A.I.) and the DFG within SFB 631 and the Nano
  Initiative Munich (NIM) (E.K., G.G. and I.C.) and NSF under Grant No. 0653417 (S.Y.).
\end{acknowledgments}

\vspace{0.5 cm}

{\bf METHODS}

\vspace{0.5 cm}

{\bf Hamiltonian of the laser driven coupled electron-nuclei system.}

The Hamiltonian of the solid-state CPT system we consider is $H_{CPT} = H_0 +
{H}_{\mathrm{laser}}  + H_{\mathrm{hyp}}$ with
\begin{equation}
 H_0 = \omega_x \sigma_{\downarrow \downarrow} + \omega_t\sigma_{tt} ,
\end{equation}
\begin{equation}
    H_{\mathrm{laser}} =  [\Omega_p
                     \sigma_{t\uparrow} e^{-i\omega_p t}
                           + \Omega_c
                    \sigma_{t\downarrow} e^{-i\omega_c t}
                           + h.c. ].
\end{equation}
Here, $\uparrow$ ($\downarrow$) denotes $\uparrow_x$ ($\downarrow_x$). The
definition of $H_{\mathrm{hyp}}$  as well as the  quantities  appearing in $H_0$
and $H_{\mathrm{laser}}$ are given in the main text.

\vspace{0.5 cm}

{\bf Rate equation description of nuclear spin dynamics.}

The semiclassical limit can be derived from the master equation by
replacing the collective spin decay by independent decay of
individual spins. This is readily accomplished by making the
substitution $I_+\rho^n I_-= \sum_{ij} g_ig_j I^i_+\rho^n I^j_+
\rightarrow \sum_{i} g_i^2 I^i_+\rho^n I^i_+ $ [and correspondingly
for other terms in Eq.(\ref{eq:D})]. We coarse grain the nuclear
motion with regard to the electron dynamics and from the new master
equation we obtain a rate equation. We introduce a shell model of
the QD with $M$ different classes of nuclear spins (Fig.~1b); the
nuclei in class ($\nu$) have identical $g_\nu$ and their net spin
polarization is $m_\nu = \frac{1}{2}(N_{\nu}^+ - N_{\nu}^-) =
\langle \sum_{i \epsilon \nu} I^i_x \rangle$, where $N_{\nu}^+$
($N_{\nu}^-$) denote the total number of up (down) spins in class
($\nu$). The derived rate equation for the joint probabilities
$\mathcal{P}(\{m_\mu\})$ associated with the nuclear spin
configuration $\{m_\mu\}$ is given by

\begin{eqnarray}\label{eq:nuc_rate}
   \frac{\partial \mathcal{P}(\{ m_\mu \})}{\partial t}
   &  \! =  \! & \sum_{\nu}^M \mathcal{P}(\{ \bar{m}_{\mu}\}) N_{\nu}^-(\{ \bar{m}_{\mu}\}) \Gamma_+^{\nu}(\{ \bar{m}_{\mu}\}) \nonumber \\
 &    +   & \sum_{\nu}^M \mathcal{P}(\{ \tilde{m}_{\mu} \})N_{\nu}^+(\{ \tilde{m}_{\mu}\})  \Gamma_-^{\nu}(\{ \tilde{m}_{\mu} \})
 \nonumber \\
 & - & \sum_\nu^M \mathcal{P}(\{ m_\mu \})[N_{\nu}^- \Gamma_+^\nu(\{ m_{\mu} \}) + N_{\nu}^+ \Gamma_-^{\nu}(\{ m_{\mu} \})] \nonumber
\end{eqnarray}
where $\Gamma_{\pm}^{\nu}(\{m_{\mu}\}) = (\frac{g
g_{\nu}}{4\omega_x})^2 \frac{\Gamma}{2} \rho_{tt}(\{m_{\mu}\})$ are
the rates at which nuclear spins of the $\nu$th class are flipped if
the nuclear spin polarizations in each class are given by
$\{m_{\mu}\}$. $\{ \bar{m}_{\mu}\}$ ($\{ \tilde{m}_{\mu} \}$)
denotes the nuclear spin configuration that differs from the
configuration $\{m_{\mu}\}$ only in the $\nu$th class, with
polarization $m_{\nu} - 1$ ($m_{\nu} +1$).

We numerically simulate the evolution of the nuclear spins with a
Monte Carlo method. We assume in our numerical simulations that the
QD contains 100 nuclear spins. We group these spins into five
concentric shells ($M=5$) with different hyperfine coupling
constants that are determined by the 3D Gaussian electronic envelope
function (Fig.~1b). The coupling constants $gg_i$ for these shells
are 0.0934$\Gamma$, 0.0828$\Gamma$, 0.0678$\Gamma$, 0.0513$\Gamma$,
0.0358$\Gamma$ and the corresponding total numbers of nuclear spins
in each shell are 2, 8, 16, 28, 46. The coupling constants are
chosen to ensure that the standard deviation of the Overhauser field
seen by the QD electron for nuclei in a completely mixed state
satisfies $\sigma_{OF}(\rho) = \frac{\Gamma}{4}$. We do not keep
track of the exact configuration within each class ($\nu$) of
nuclear spins and assume that any configuration of spins leading to
the same $m_{\mu}$ is equally likely and that the nuclear spin
distribution in each shell is independent of the other shells.

\vspace{0.5 cm}

{\bf Nuclear spin flip rates for homogeneous hyperfine coupling}

For any eigenstate $| m \rangle$ of $I_x$ with $I_x \ket{m} = m/\sqrt{N}\ket{m}$
($\Gamma_\mathrm{nuc}\ket{m} = \Gamma^m_\mathrm{nuc}\ket{m}$) the nuclear spin
flip rate in negative (positive) direction is given by $D^- = \langle I_+I_-
\Gamma_\mathrm{nuc} \rangle_m$ ($D^+ = \langle I_-I_+ \Gamma_\mathrm{nuc}
\rangle_m$). In the semiclassical limit under the assumption $\langle
I^i_+I^j_-\rangle=0$ ($i\neq j$) the rates are simply given as $D_{sc}^\mp=
(\frac{1}{2}\pm\frac{m}{N})\Gamma_\mathrm{nuc}^m$. For the quantum description the
characterization via the spin projection quantum number $m$ is not sufficient; the
rates also depend on the symmetry of the nuclear state, quantified by the total
spin $J\in \{0,...,N/2\}$. For a Dicke state $\ket{J,m}$ the rates are given as
$D_{qm}^\mp = \langle I_\pm I_\mp
\Gamma_\mathrm{nuc}\rangle_{J,m}=\frac{1}{N}[J(J+1)-m(m\mp1)]\Gamma_\mathrm{nuc}^m$.
For the statistically relevant $J$-subspaces [$J =\mathcal{O}( \sqrt{N})$] we find
that $D_{qm}^\mp$ are in good agreement with those obtained in the semiclassical
limit for small $m$ values.

\vspace{0.5 cm}

{\bf Generalized Overhauser field}

For the main part of the nuclear Hilbert space -- namely the domain
where the operator $I_x + \delta/g$ is large (recycling region) --
the $\epsilon$ correction to the Overhauser field represents a
negligible perturbation to the hyperfine interaction. However, in
the domain of small eigenvalues of $I_x + \delta/g$ (trapping
region) this perturbative picture is not trivially justified.
However, for the diffusive dynamics we are mainly interested in the
number of eigenstates in a region $L=(-\delta-\eta,-\delta+\eta)/g$
around $-\delta/g$. For $I_x$ the number of eigenstates with small
eigenvalues is very large (exponential in the number of spins),
which is favorable for our scheme. For the generalized Overhauser
field it is given by
$D(\eta,\epsilon)=\int_{-\delta-\eta}^{-\delta+\eta}dE
\mathrm{~Tr}(\delta(E-I_x-\epsilon I_+I_-))$ and deviates from the
number of unperturbed eigenstates:
\begin{align}
D(\eta,\epsilon)=D(\eta,0) + \sum_{n=1}^\infty U^{(n)}.
\end{align}
The sum $ \sum_{n=1}^\infty U^{(n)}$ can be upper bounded by $\sim
D(\eta,0)$ for large trapping regions $\eta \gg \epsilon$, i.e. the
number of eigenstates changes at most by a factor of order 1.
Numerical calculations for $N=10^4$ inhomogeneously coupled spins
show that even for $\eta \sim \epsilon$ the number of states in both
the perturbed and unperturbed case differ only by a few percent.

\newpage

\noindent {\bf Figure Captions}

\vspace{0.5 cm}

\noindent {\textbf{Figure 1 The energy level diagram of a
solid-state emitter.}} (a) The electron spin state
$|\uparrow_x\rangle$ ($|\uparrow_x\rangle$) is resonantly coupled to
a trion state with an $x$ ($y$) polarized laser field with Rabi
frequency $\Omega_p$ ($\Omega_c$). In Voigt geometry the oscillator
strengths of the two transitions are identical, leading to
spontaneous emission rates with equal strength $\Gamma/2$. Optical
excitation allows for energy conserving hyperfine flip-flop
transitions that result in nuclear spin diffusion; these
second-order processes are depicted using dashed curves.
  (b) The confined electron wave-function leads to inhomogeneous hyperfine coupling with the nuclei. In the
  simulations, we assume that the dot
  can be described as consisting of 5 different classes  of nuclei.
  All nuclei within a class have identical hyperfine coupling,
  with strength determined by the electron wave function.

\vspace{0.5 cm}

\noindent {\textbf{Figure 2 Nuclear spin state selective coherent
population trapping.}} (a) The absorption lineshape in the presence
of hyperfine interactions with quantum dot nuclei for Rabi
frequencies $\Omega_p = \Omega_c = 0.2~\Gamma$ (red) and $\Omega_p =
\Omega_c = 0.4~\Gamma$ (blue): in stark contrast to the standard
coherent population trapping profile (dashed lines), the dark
resonance is drastically broadened (solid lines).  The broadening of
the dark resonance is a consequence of the fact that optical
excitation induced nuclear spin diffusion allows the coupled
electron-nuclei system to find a Overhauser field configuration that
satisfies the dark state condition for a broad range of initial
laser detunings.
 (b) The standard deviation of
the Overhauser field $\sigma_{OF}$ for  $\Omega_p = \Omega_c =
0.2~\Gamma$ (red line) and $\Omega_p = \Omega_c = 0.4~\Gamma$ (blue
line)) is reduced to the level below that of a single nuclear spin
flip (green dashed line). The dashed black line shows the standard
deviation in the absence of laser drive. (c) The final Overhauser
field distribution could be determined by using a fast scan of the
probe laser across the optical resonance. The solid (blue) curve
shows the absorption lineshape obtained when the system is initially
prepared in a dark state by setting $\Delta \omega_p = -0.2 \Gamma$:
the width and the depth of the dark resonance reveals information
about the Overhauser field distribution. If the experiment is
carried out by starting out in a random nuclear spin state, the
observed lineshape is close to a Lorentzian (dashed green curve).

\vspace{0.5 cm}

\noindent {\textbf{Figure 3 Homogeneously coupled electron-nuclei
system.}} (a) Nuclear spin diffusion rates depending on the nuclear
spin projection  $m$ assuming homogeneous coupling (cf.
Eq.~(\ref{eq:end})). Parameters are $N=4\times 10^{4}$,
$\Gamma=1$~GHz, $A=\omega_x=100$~$\mu$eV and $\Omega=0.1$~GHz. The
quantum mechanical rates are taken for the subspace $J=\sqrt{N/2}$.
(b) The dependence of the steady state value of the nuclear
Overhauser field standard deviation as a function of $\Omega$,
calculated using the fully quantum model. The increase in standard
deviation for $\Omega_p = \Omega_c < 0.2 \Gamma$ is a consequence of
the fact that the coupled electron-nuclei system spends a
substantial amount of time outside the transparency region. The
inset shows the steady state population of $I_x$ eigenstates for
$\Omega_p=\Omega_c =0.02 \Gamma$. Despite the finite standard
deviation $\sigma \approx 9$ the system is strongly peaked around
$m=0$.

\newpage

 \includegraphics[width=0.45\textwidth]{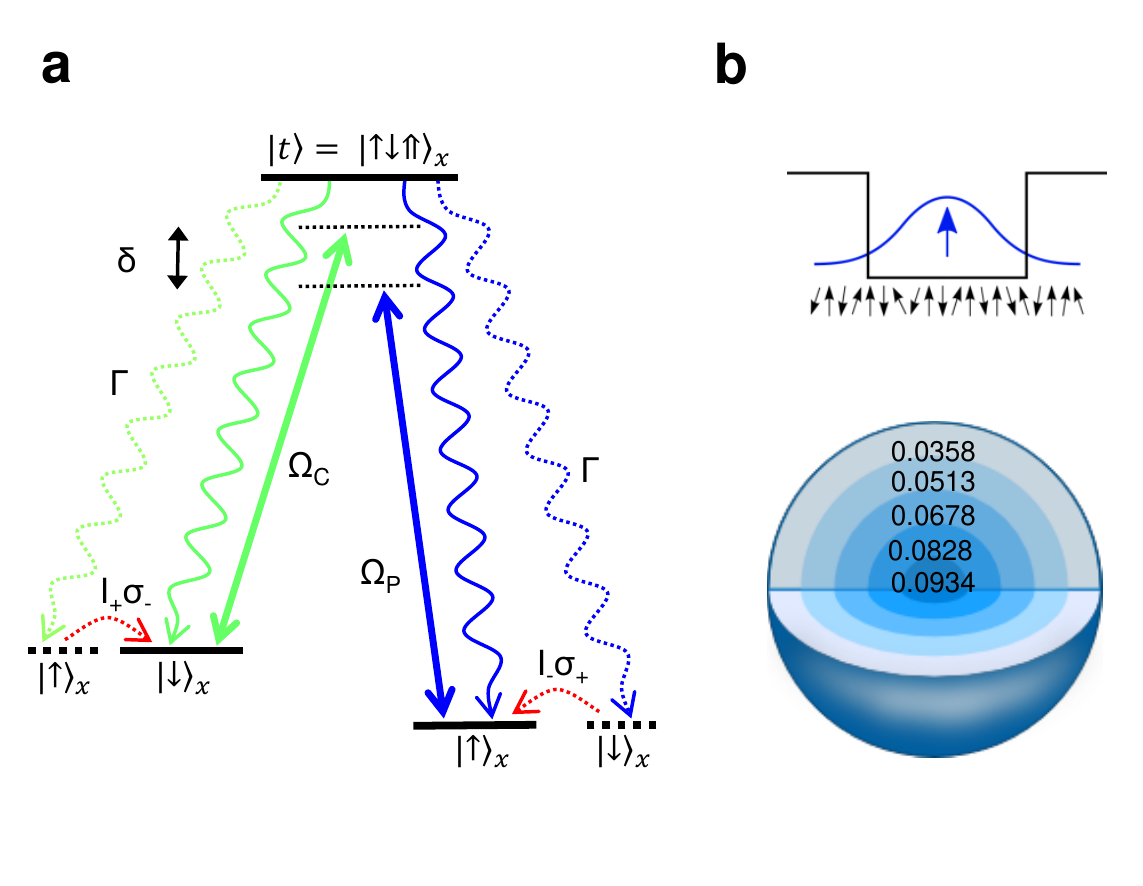}

\newpage

.

\newpage

 \includegraphics[width=0.45\textwidth]{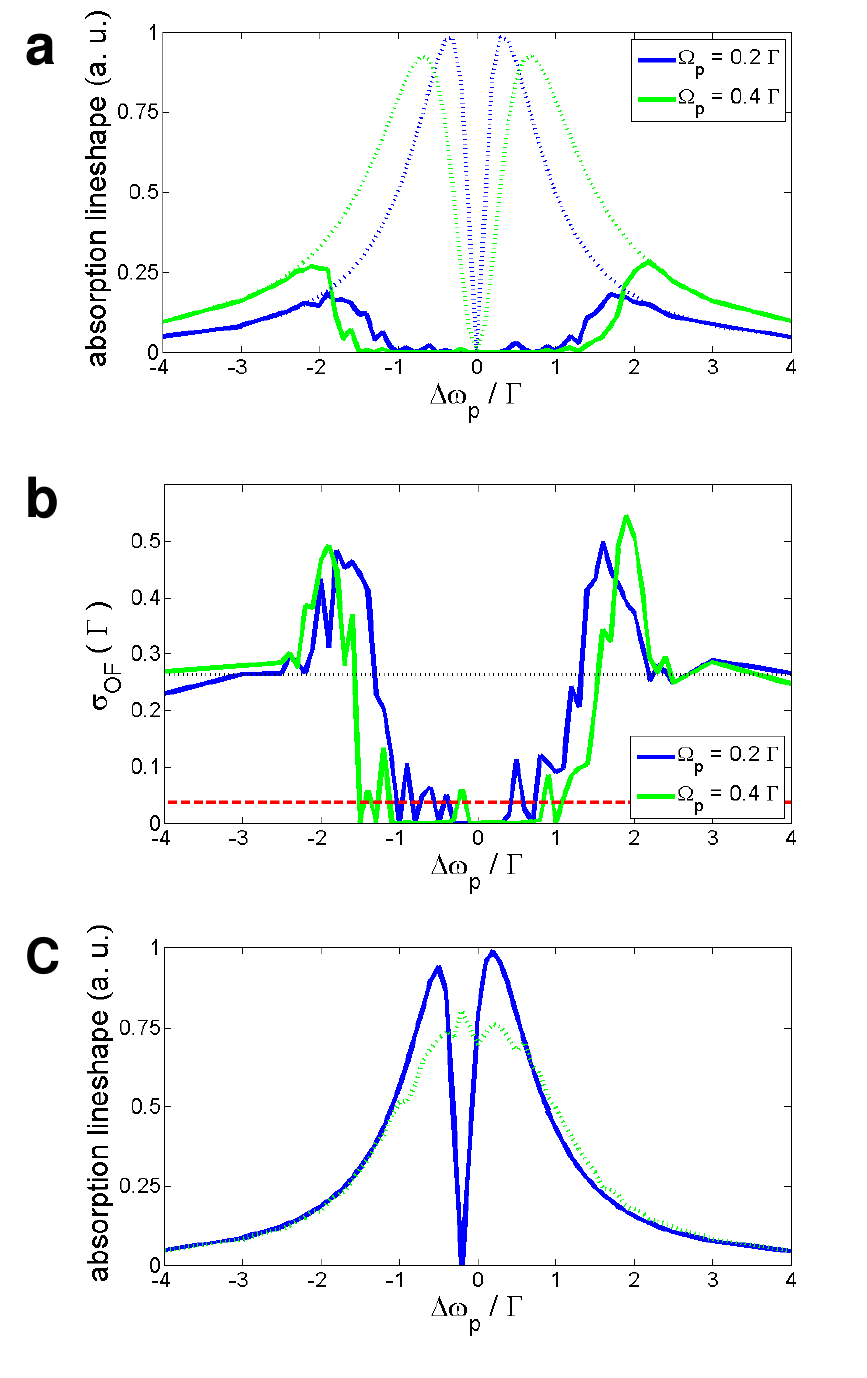}

\newpage

.
\newpage

 \includegraphics[width=0.45\textwidth]{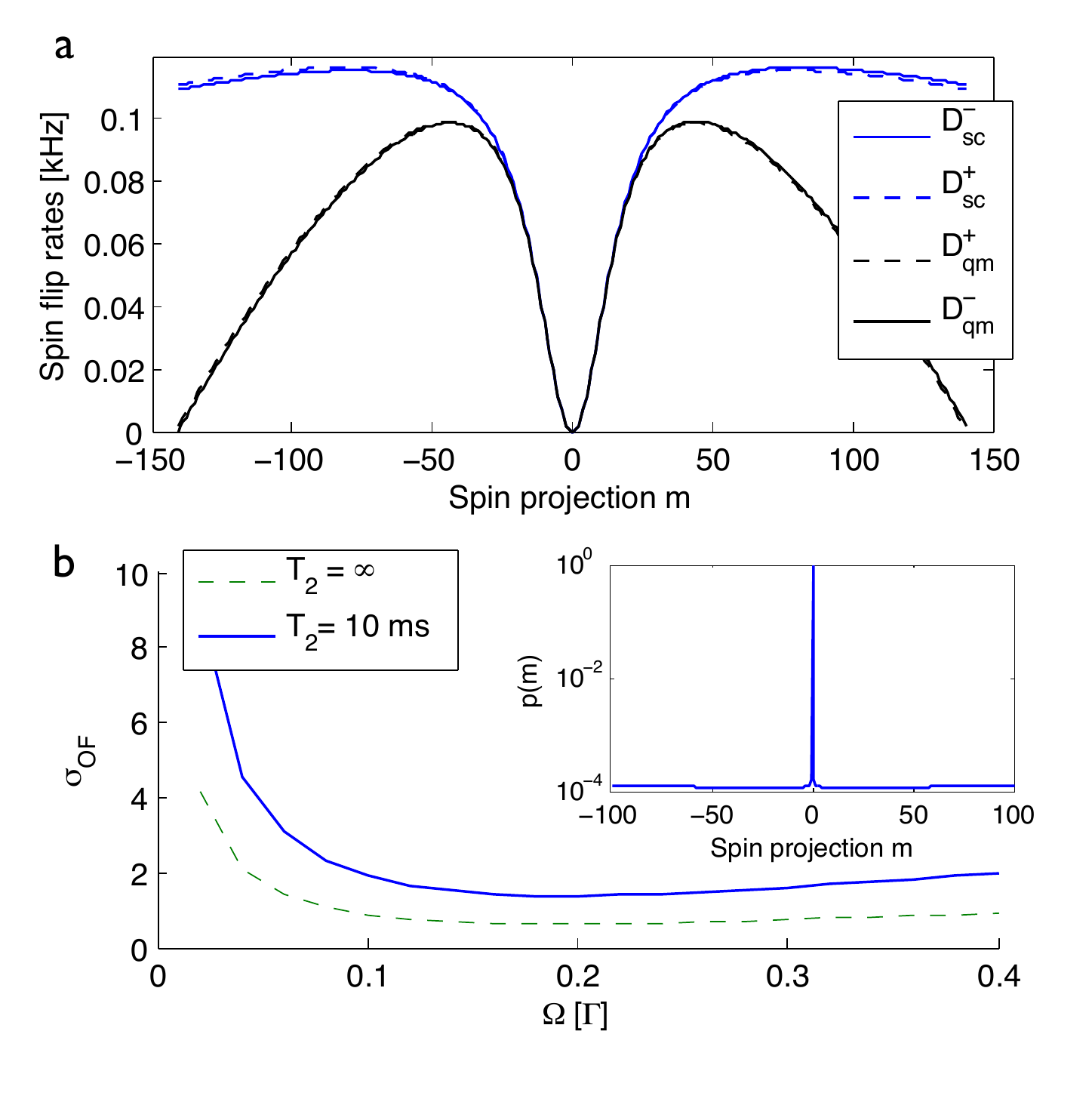}


\begin{thebibliography}{99}
\bibitem{SW-note} Contribution to the master equation arising from the
Schrieffer--Wolff transformation applied to the laser coupling terms
do not lead to terms of order $\epsilon$, provided that the laser
polarizations match that of the corresponding optical transitions.
\bibitem{densitymatrix} For Eq.~(\ref{eq:fin}) to generate a physical
(completely positive) dynamics, $I_x$ must commute with $\rho$ at
all times. This is ensured by Eq.~(\ref{eq:fin}) (in the homogeneous
case) provided it holds initially, (e.g. for initially fully mixed
nuclear spins). We also note that since $[I_x,I_+I_-]=0$ in the
homogeneous case, inclusion of the generalized Overhauser field
$\tilde{I}_x=I_x+\epsilon I_+I_-$ is straightforward.
\bibitem{off-resonant} An example of such a process is off-resonant
hyperfine-assisted laser scattering $ \propto \epsilon^2
(\Gamma/\omega_x)^2$.
\bibitem{scat-rate}  This assumption is justified by the fact that the degeneracy of the
the available Overhauser field configuration drops for larger
$\langle I_x \rangle > A_H/\sqrt{N}$. In addition, the random walk
of the Overhauser field is biased since the rates of flipping a spin
up or down come with a factor proportional to the number of
available final configurations. This bias becomes large for $\langle
I_x \rangle > A_H/\sqrt{N}$ and prevents the Overhauser from
exploring extreme polarizations.
\bibitem{Bayer-desc} Nuclear spin preparation in these experiments could be
considered in the frame of a {\sl time-dependent dark-state} where
the electron spin is in a superposition state with a time-dependent
phase arising from the effective Zeeman splitting. This phase
evolves in a way to ensure that at the arrival time of the laser
pulse, the electron is in a dark superposition of the spin states;
this condition is enforced by nuclear spin polarization that is
different for each QD.
\end{thebibliography}
\end{document}